\documentclass[12pt,preprint]{aastex}
\usepackage{graphicx,natbib,amsmath}

\newcommand{\lapprox} {\, \lower3pt\hbox{$\sim$}\llap{\raise2pt\hbox{$<$}}\,}
\newcommand{\gapprox} {\, \lower3pt\hbox{$\sim$}\llap{\raise2pt\hbox{$>$}}\,}

\begin{document}

\title{The Formation of Kappa-Distribution \\ Accelerated Electron Populations in Solar Flares}

\author{Nicolas H. Bian\altaffilmark{1},
        A. Gordon Emslie\altaffilmark{2},
        Duncan J. Stackhouse\altaffilmark{1}, and
        Eduard P. Kontar\altaffilmark{1} }

\altaffiltext{1}{School of Physics \& Astronomy, University of Glasgow, Glasgow G12 8QQ, Scotland, UK \\
(n.bian@physics.gla.ac.uk, d.stackhouse.1@research.gla.ac.uk, eduard@astro.gla.ac.uk)}

\altaffiltext{2}{Department of Physics \& Astronomy, Western Kentucky University, Bowling Green, KY 42101 (emslieg@wku.edu)}

\begin{abstract}

Driven by recent {\em RHESSI} observations of confined loop-top hard X-ray sources in solar flares, we consider stochastic acceleration of electrons in the presence of Coulomb collisions. If electron escape from the acceleration region can be neglected, the electron distribution function is determined by a balance between diffusive acceleration and collisions.  Such a scenario admits a stationary solution for the electron distribution function that takes the form
of a kappa distribution. We show that the evolution toward this kappa distribution involves a ``wave front'' propagating forwards in velocity space, so that electrons of higher energy are accelerated later; the acceleration time scales with energy according to $\tau_{\rm acc} \sim E^{3/2}$. At sufficiently high energies escape from the finite-length acceleration region will eventually dominate. For such energies, the electron velocity distribution function is obtained by solving a time-dependent Fokker-Planck equation in the ``leaky-box'' approximation. Solutions are obtained in the limit of a small escape rate from an acceleration region that can effectively be considered a thick target.

\end{abstract}

\keywords{acceleration -- Sun: activity -- Sun: flares -- Sun: X-rays, gamma rays}

\section{INTRODUCTION}

Ever since the first hard X-ray observations of solar flares \citep{1959JGR....64..697P}, it has been realized that these events  are responsible for the acceleration of copious amounts of charged particles, in particular deka-keV electrons.  Fifty years of observations have revealed considerable insight into the spectral, temporal, and spatial properties of these accelerated electrons; however, the underlying mechanism responsible for their acceleration remains largely undetermined.  A major objective of contemporary high-energy solar physics research is, then, to understand not only the {\it propagation} of  accelerated electrons within the source but also the physics of their {\it acceleration}. To do this requires that we obtain information on the hard X-ray emission produced by accelerated electrons with spectral, spatial and temporal resolutions sufficiently precise to probe the emergence of the accelerated electron spectrum from the initial quasi-Maxwellian population.  Acquisition of such data was a key element in the design of the {\em RHESSI} instrument \citep{2002SoPh..210....3L}.

Proposed acceleration mechanisms include acceleration by large-scale coherent sub-Dreicer electric fields \citep[e.g.,][]{1994ApJ...435..469B} and by supra-Dreicer electric fields in thin reconnecting current sheets \citep[e.g.,][]{1993SoPh..146..127L,1996ApJ...462..997L}. However, these models face serious challenges in terms of the properties of the source (e.g., fine fragmentation, efficient pitch-angle scattering) in order to avoid unacceptably large unidirectional currents \citep{1985ApJ...293..584H,1995ApJ...446..371E}. Further, there is growing observational evidence \citep[e.g.,][]{1980ApJ...239L..85K,2006ApJ...653L.149K} that the overall accelerated electron distribution has an angular distribution that is nearly isotropic.  Combined, these theoretical and observational considerations favor a stochastic acceleration model invoking plasma turbulence where particles undergo multiple energetic ``boosts'' by an ensemble of scattering centers \citep[e.g.,][for reviews]{1994ApJS...90..623M, 1997JGR...10214631M, 2012SSRv..173..535P, 2012ApJ...754..103B}. Stochastic acceleration models have been applied to solar flares \citep[e.g.,][]{PhysRev.111.1206,1979AIPC...56..135R,1996ApJ...461..445M,2010ApJ...712L.131P}, and often share the property that the acceleration can be described by a second-order velocity diffusion coefficient $D_{vv}$.

We therefore here consider a model that involves acceleration by a stochastic process, modeled through a diffusion term in the Fokker-Planck equation describing the evolution of the electron phase-space distribution function, coupled with particle transport that consists of two components: {\it in situ} Coulomb collisions with the background plasma, and escape associated with the finite length of the acceleration region. In general, the results are characterized by four governing timescales: the acceleration timescale $\tau_{\rm acc}$, the collisional deceleration timescale $\tau_{\rm c}$, the collisional diffusion timescale $\tau_{\rm d}$, and the escape timescale $\tau_{\rm esc}$. In the region in velocity space where escape can be neglected, the electron distribution function is driven toward a steady state corresponding to a balance between diffusive acceleration and collisional energy losses. For a velocity diffusion coefficient $D_{vv} \sim 1/v$, this equilibrium state takes the form of a kappa distribution, which transitions smoothly from a Maxwellian low-energy core to a power-law high-energy tail \citep{0038-5670-9-3-R04,1977ApJ...211..270B, 1985PhRvL..54.2608H,1998GeoRL..25.4099M,2004PhPl...11.1308L}.
This result is encouraging, since kappa distributions have been used to characterize particle distribution functions in a variety of space plasma scenarios \citep[e.g.,][]{2009JGRA..11411105L},
including electrons in solar flares \citep{2009A&A...497L..13K,2013ApJ...764....6O}

The context and general properties of the acceleration model are detailed in Section~\ref{general}. In Section~\ref{kap}, we consider the steady-state solution for the accelerated electron distribution in the case of an acceleration model characterized by a diffusion coefficient with an inverse dependence on velocity.  This takes the form of a kappa distribution, which is characterized by two parameters, one of which is a characteristic velocity scale (e.g., the thermal velocity associated with the Maxwellian core) and the other is the dimensionless ratio of the acceleration time to the collisional deceleration time.  The dimensionless parameter, denoted by $\kappa$, is simply related to the power-law spectral index $\delta$ of the electron energy flux, and hence to $\gamma$, the power-law index of the emitted hard X-ray bremsstrahlung spectrum.  In Section~\ref{evol} we characterize the time evolution toward the asymptotic kappa distribution as an advancing wavefront in velocity space, and we find that the acceleration of electrons to energy $E$ occurs on a timescale $\tau_{\rm acc} \propto E^{3/2}$.

At sufficiently high energies, particle escape associated with the finite length of the acceleration region can modify this asymptotic form, and we consider this effect in Section~\ref{escape-analysis}. In Section~\ref{numerical}, we present numerical solutions of the basic Fokker-Planck equation and we discuss the extent to which the numerical results confirm the analytic results of the previous sections. In Section~\ref{stationary} we analyze a model that is appropriate to acceleration by a coherent large-scale electric field, showing that the presence of efficient pitch-angle scattering can lead to isotropization of the distribution function and hence can produce an effect akin to stochastic acceleration over a wide velocity range. We determine the conditions for the turbulent diffusion coefficient in such a model to take the desired form $D_{vv} \sim 1/v$ and we discuss the constraints that the model imposes on the magnitude of the electric field. In Section~\ref{conclusion} we summarize the results obtained.

Overall, our results lead to the characterization, over a wide velocity range, of the evolution of the electron distribution toward its asymptotic form, an analysis pertinent to the study of a variety of stochastic acceleration models that have been associated with hard X-ray production during the impulsive phase of a solar flare.

\section{CONTEXT AND GENERAL DESCRIPTION OF THE ACCELERATION MODEL}\label{general}

{\em RHESSI} has revealed \citep[e.g.,][]{2008ApJ...673..576X} the presence of coronal hard X-ray flare sources with a background density sufficiently high that the accelerated electrons are collisionally stopped in the corona, rather than streaming through it and impacting on the chromosphere to produce hard X-ray footpoints \citep[cf.][]{2003ApJ...595L.107E}. The spatial distribution of these thick-target coronal hard-X ray sources exhibits a core region where acceleration occurs, surrounded by a halo where escaping high-energy electrons are collisionally stopped \citep{2008ApJ...673..576X}.  Since the acceleration and hard X-ray emitting regions are coincident, these flares have opened new horizons for the study of acceleration processes, inasmuch as they permit determination of the length of, and density within, the acceleration region \citep{2008ApJ...673..576X,2011ApJ...730L..22K,2012A&A...543A..53G}, and hence the number of particles available for acceleration and the specific acceleration rate (electrons~s$^{-1}$ per ambient electron), a quantity that measures the efficiency of the acceleration process \citep{2013ApJ...766...28G}. Spectroscopic imaging observations with {\em RHESSI} also suggest the presence of turbulence (due to, e.g., fluctuations in the magnetic field) in these coronal loops \citep{2011ApJ...730L..22K}, resulting in both pitch-angle scattering \citep{2014ApJ...780..176K} and cross-field transport \citep{2011A&A...535A..18B} of high-energy electrons.

Our aim is to develop a model for the electron phase-space distribution function in coronal thick-target sources, using a Fokker-Planck equation that includes the combined effects of turbulent acceleration and Coulomb collisions with the dense background plasma:

\begin{equation}\label{fund}
\frac{\partial f}{\partial t}=\frac{1}{v^{2}} \, \frac{\partial }{\partial v} \,
\left \{ v^{2} \left [ \left ( \frac{\Gamma \, v_{\rm te}^{2}}{2v^{3}}+D_{\rm turb}(v) \right ) \,
\frac{\partial f}{\partial v}+\frac{\Gamma}{v^{2}} \, f \right ] \right \} \,\,\, .
\end{equation}
Here $f$ (electrons~cm$^{-3}$~(cm~s$^{-1}$)$^{-3}$) is the phase-space distribution function of electrons, averaged over the acceleration region volume, and we use a simplified form of the collision operator applicable to the solar flare situation \citep[cf.][]{2014ApJ...787...86J}, in which the background electrons are modelled as a heat bath at a fixed temperature $T$(K).  In Equation~(\ref{fund})

\begin{equation}\label{vte-def}
v_{\rm te}=\sqrt{2k_{B}T/m_{e}}
\end{equation}
is the thermal speed (with $k_B$ (erg~K$^{-1}$) the Boltzmann constant and $m_e$ (g) the electron mass),

\begin{equation}\label{gamma-def}
\Gamma = \frac {4\pi e^{4} \ln\Lambda \, n} {m_{e}^{2}}
\end{equation}
is the collision parameter (with $n$ (cm$^{-3}$) the density of background electrons, $e$ (esu) the electronic charge, and $\ln \Lambda$ the Coulomb logarithm), and $D_{\rm turb}(v) \equiv D_{vv}$ (cm$^2$~s$^{-3}$) is the diffusion coefficient in velocity space associated with an as yet unspecified stochastic acceleration mechanism.

There are three characteristic timescales in the stochastic acceleration model represented  by Equation~(\ref{fund}), viz.

\begin{itemize}

\item the {\it acceleration time} $\tau_{\rm acc}$, defined through

\begin{equation}\label{tacc-def}
\frac{1}{v^{2}} \, \frac{\partial }{\partial v} \, \left \{ v^{2} \left [ D_{\rm turb}(v) \,
\frac{\partial f}{\partial v} \right ] \right \} \simeq \frac{f}{\tau_{\rm acc}(v)} \,\,\, ; \,\,\, \tau_{\rm acc}(v)=\frac{v^2}{D_{\rm turb}(v)} \,\, \, ;
\end{equation}

\item the {\it collisional deceleration/friction time} $\tau_{\rm c}$, defined through

\begin{equation}\label{tc-def}
\frac{\Gamma}{v^{2}} \, \frac{\partial f}{\partial v} \,\,\, \simeq \frac{f}{\tau_{\rm c}(v)} \,\,\, ; \,\,\,
\tau_{\rm c}(v) \simeq \frac{v^3} {\Gamma} \,\,\, ; \qquad {\rm and}
\end{equation}

\item the {\it collisional diffusion time} $\tau_{\rm d}$, defined through

\begin{equation}\label{tdiff-def}
\frac{1}{v^{2}} \, \frac{\partial }{\partial v} \, \left \{ v^{2} \left [ \frac{\Gamma v_{te}^{2}}{2v^{3}} \,
\frac{\partial f}{\partial v} \right ] \right \} \simeq \frac{f}{\tau_{\rm d}(v)} \, \, \, ; \,\,\, \tau_{\rm d}(v) \simeq
\frac{2 v^{5}}{\Gamma v_{te}^{2}} \,\,\, .
\end{equation}

\end{itemize}

Many important properties of the model are more conveniently derived by recasting the Fokker-Planck Equation~(\ref{fund}) in the form studied by \citet{2005PhRvE..72f1106C} and \citet{2010PhyA..389.1021L}:

\begin{equation}\label{fp-transformed}
\frac{\partial f}{\partial t}=\frac{1}{{v^{2}}} \, \frac{\partial }{\partial v} \,
\left [ v^{2} \, D(v) \left ( \frac{\partial f}{\partial v} +f \, U^\prime(v) \right ) \right ] \,\,\, ,
\end{equation}
with

\begin{equation}\label{dv}
D(v)=\frac{\Gamma  \, v_{\rm te}^{2}}{2  \, v^{3}} + D_{\rm turb}(v)
\end{equation}
and

\begin{equation}\label{uprime}
U^\prime(v)=\frac{\Gamma}{v^{2}D(v)} =  \left ( {v_{\rm te}^2 \over 2v} + { v^2 \, D_{\rm turb}(v) \over \Gamma } \right )^{-1} \,\,\, .
\end{equation}

Observations of quantities related to $f(v,t)$ are generally averaged over the pertinent instrument time resolution. Specifically, imaging spectroscopy hard X-ray observations from {\em RHESSI} \citep[see][for a review]{2011SSRv..159..301K} are limited to the time it takes to develop a full set of spatial Fourier components of the source; this takes a full spacecraft rotation period of several seconds.  Given that the timescales for acceleration, collisional energy loss, and escape are, for typical conditions in loop-top coronal hard X-ray sources \citep{2013ApJ...766...28G}, less than a second, it follows that a quasi-steady-state scenario is of considerable relevance and interest.  The stationary ($\partial/\partial t = 0$) solution of the Fokker-Planck equation~(\ref{fp-transformed}) is

\begin{equation}\label{ss}
f(v)=A \, e^{-U(v)} \,\,\, ,
\end{equation}
where $A$ is a normalization constant. This result is sufficiently general to permit the determination of the steady-state distribution $f(v)$ of energetic electrons in a collisional plasma, given a specific choice of the turbulent velocity-space diffusion coefficient $D_{\rm turb}(v)$ or, equivalently, the function $U(v)$.

\section{THE KAPPA DISTRIBUTION AS A STATIONARY SOLUTION}\label{kap}

The stationary state (\ref{ss}) has been written in the form of a Gibbs-Boltzmann distribution with the function $U(v)$ playing the role of a potential. It is well known that such distributions globally minimize the Helmoltz free-energy functional $F[f]=E[f]-S[f]$ where $E[f]=\int U f \, d\mathbf{v}$ is the potential energy and $S[f]=-\int f \ln f \, d\mathbf{v}$ is the Boltzmann entropy. This property of the equilibrium state (\ref{ss}) is intimately related to the existence of a variational principle underlying the Fokker-Planck equation (\ref{fp-transformed}), which dictates that the time-dependent solution $f(v,t)$ evolves according to the following constraint \citep{2005PhRvE..72f1106C} on the functional $F[f(v,t)]$:

\begin{equation}\label{cst}
\dot{F}=-\int \frac{D(v)}{f} \, \left ( \frac{\partial f}{\partial \mathbf{v}}+f \, \frac{\partial U}{\partial \mathbf{v}} \right )^{2}d\mathbf{v} \, \leq \, 0 \,\,\, .
\end{equation}
Therefore, if $F$ is bounded from below, the distribution function converges toward the stationary state (\ref{ss}) as $t\rightarrow \infty$, corresponding to a statistical equilibrium between diffusive acceleration and collisional drag.  Indeed, the electron distribution function will steadily converge toward the stationary state~(\ref{ss}) provided the zero-flux boundary condition $v^{2} \, D(v) (\partial f/\partial v +f \, U^\prime(v) ) \rightarrow 0$ as $v \rightarrow \infty$.

When $D_{\rm turb}=0$, Equation~(\ref{uprime}) shows that the potential $U(v)$ is quadratic in $v$, so that, by Equation~(\ref{ss}), a steady-state Maxwellian distribution is obtained. In this case the Fokker-Planck equation describes collisional relaxation toward thermal equilibrium at temperature $T$. However, Equation~(\ref{uprime}) also shows that a steady-state Maxwellian distribution of electrons can be achieved in the presence of a finite level of turbulence $D_{\rm turb}\neq 0$ provided $D_{\rm turb} \sim 1/v^3$, in which case the acceleration time given by Equation~(\ref{tacc-def}) obeys a velocity dependence\footnote{acceleration times $\tau_{\rm acc} \sim v^{5}$, corresponding to $D_{\rm turb}\sim v^{-3}$, are produced by Gaussian isotropic spectra of electrostatic fluctuations, not necessarily thermal, in the plasma \citep[see][]{1992PhRvL..69.1831R}} identical to that of collisional diffusion $\tau_{acc}(v)\sim \tau_{d}(v) \sim v^{5}$. Therefore, the presence of a Maxwellian distribution of plasma electrons is \emph{not} synonymous with a state of thermal equilibrium.  This fact may complicate the interpretation of spectroscopic data, inasmuch as the temperature inferred from the shape of the electron distribution function may also include a turbulent broadening component \citep[e.g.,][]{1986ApJ...301..975A}.

The distribution function of deka-keV electrons in solar flares is generally well described by a Maxwellian core with a power-law high-energy tail \citep[see, e.g.,][]{2003ApJ...595L..97H}. In an attempt to account for this behavior, let us consider a turbulent diffusion coefficient of the form

\begin{equation}\label{dturb-result}
D_{\rm turb} (v) = {D_{0} \over v} \,\,\ ,
\end{equation}
from which it follows (Equation~(\ref{tacc-def})) that the acceleration time, defined as $\tau_{\rm acc} (v) \equiv {v^2 / D(v)}$, is given by

\begin{equation}\label{tau-acc-v3d0}
\tau_{\rm acc} (v) = {v^3 \over D_0} \,\,\, .
\end{equation}
For this case, the acceleration time $\tau_{\rm acc}$ and the collisional deceleration time $\tau_{\rm c}$ have the same velocity dependence, $\tau_{\rm acc}(v) \propto \tau_{\rm c}(v) \propto v^3$.  Therefore we can define the dimensionless constant

\begin{equation}\label{kappadef}
\kappa = {\tau_{\rm acc}(v) \over 2 \, \tau_{\rm c}(v)} = \frac{\Gamma}{2 D_{0}} \,\,\, ,
\end{equation}
(the reason for the factor 2 will be evident shortly).  With this identification, Equation~(\ref{uprime}) becomes

\begin{equation}\label{uprime-1-over-v}
U'(v) =
{2 v \over v_{\rm te}^2} \left ( 1 + {v^2 \over \kappa \, v_{\rm te}^2 } \right )^{-1} \,\,\, ,
\end{equation}
with the following solution for the potential $U(v)$:

\begin{equation}\label{u-solution}
U(v) = \kappa \, \ln \left ( 1 + {v^2 \over \kappa \, v_{\rm te}^2} \right ) \,\,\, .
\end{equation}
Thus, by Equation~(\ref{ss}), the (normalized) stationary solution is the well-known \citep[see, e.g.,][]{1968JGR....73.2839V,2009A&A...497L..13K,2009JGRA..11411105L,2013ApJ...764....6O} kappa distribution

\begin{equation}\label{kappa-dist}
f_\kappa(v)=
\frac{n_{\kappa}}{\pi^{3/2} \, v_{\rm te}^{3} \, \kappa^{3/2}} \, \frac{\Gamma(\kappa)}{\Gamma \left (\kappa-\frac{3}{2} \right ) } \,
\left ( 1+\frac{v^{2}}{\kappa \, v_{\rm te}^{2}} \right )^{-\kappa} \,\,\, ,
\end{equation}
where $n_\kappa = \int f_{\kappa}(v) \, d^{3}v$ is the number density associated with the accelerated electron distribution.

Equation~(\ref{kappa-dist}) defines a {\it kappa distribution of the first kind}, in the terminology of \citet[][their Equation~(9)]{2009JGRA..11411105L}.  We note that other authors \citep[e.g.,][]{2009A&A...497L..13K,2013ApJ...764....6O} have used a kappa distribution of the {\it second} kind \citep[again in the terminology of][their Equation~(10)]{2009JGRA..11411105L}:

\begin{equation}\label{kappa-dist-second-kind}
f_{\tilde \kappa}(v)=
\frac{n_{\tilde \kappa}}{\pi^{3/2} \, \theta^{3} \, {\tilde \kappa}^{3/2}} \, \frac{\Gamma({\tilde \kappa}+1)}{\Gamma \left ({\tilde \kappa}-\frac{1}{2} \right ) } \,
\left ( 1+\frac{v^{2}}{{\tilde \kappa} \, \theta^2} \right )^{-({\tilde \kappa}+1)}
\end{equation}
to describe the electron distribution function in solar flares.  Such authors have also used the concept of {\it kinetic temperature} $T_K$, defined such that the average energy of the electrons in the kappa distribution (\ref{kappa-dist-second-kind}) is ${\overline E} = (3/2) \, k_{B} \, T_K$ \citep[see, e.g.,][]{2013ApJ...764....6O}.  It follows that

\begin{equation}\label{kinetic_temperature}
k_{B}T_K={1 \over 2} \, m_e \, \theta^2 \left [{{\tilde \kappa} \over ({\tilde \kappa} - 3/2)}\right ]
\end{equation}
and it should be noted that the kinetic temperature $T_K$ is not to be confused with $T$ (Equation~(\ref{vte-def})), the temperature of the background Maxwellian with which the accelerated electrons interact.

It is important to note that Equations~(\ref{kappa-dist}) and~(\ref{kappa-dist-second-kind}) refer to an identical family of two-parameter distributions; only the parametric labelling of the mathematical form is different in the two descriptions. Indeed, as noted by \citet{2009JGRA..11411105L}, the changes of variable

\begin{equation}\label{change_of_paramters}
{\tilde \kappa} = \kappa - 1 ; \qquad \theta = \sqrt{\kappa \over \kappa - 1} \,\, v_{\rm te}
\end{equation}
transform Equation~(\ref{kappa-dist}) into Equation~(\ref{kappa-dist-second-kind}) {\it exactly}. \citet{2009JGRA..11411105L} note that ``the first kind of kappa distribution is less widely used than the second kind.'' However, in our ``first kind'' parametrization~(\ref{kappa-dist}), the quantity $\kappa$ has an immediate physical significance, namely the dimensionless ratio (Equation~(\ref{kappadef})) of two physical quantities: the stochastic acceleration time $\tau_{\rm acc}$ (Equation~(\ref{tacc-def})) and the collisional deceleration time $\tau_{\rm c}$ (Equation~(\ref{tc-def})) or, equivalently, the collisional parameter $\Gamma$ (Equation~(\ref{gamma-def})) and the diffusion parameter $D_0$ (Equation~(\ref{dturb-result})). We therefore submit that the form~(\ref{kappa-dist}) is a more natural choice of kappa distribution parametrization.

\begin{figure}[htpb]
\centering
\includegraphics[width=10cm]{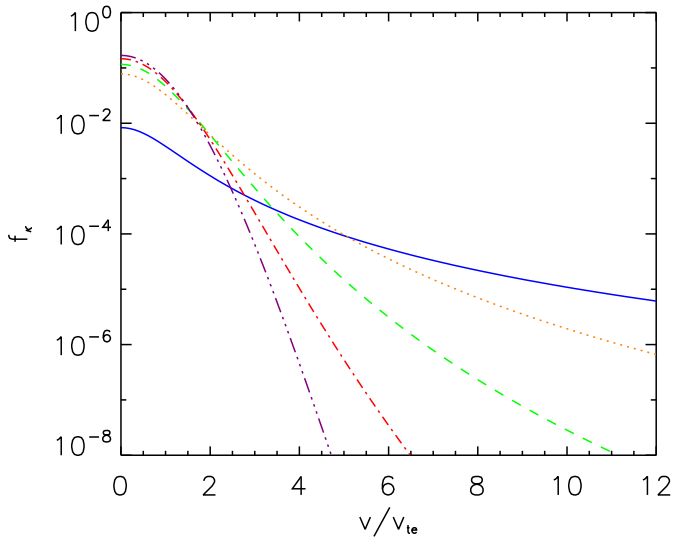}
\caption{The stationary solution kappa distribution, $f_{\kappa}$, for different values of $\kappa$, all normalized to a density $n_\kappa=1$. {\it Solid blue line:} $\kappa = 1.6$, {\it dotted orange line:} $\kappa = 3$, {\it dashed green line:} $\kappa = 5$, {\it dot-dashed red line:} $\kappa = 10$, {\it dot-dot-dot-dashed purple line:} $\kappa = 30$.  For small values of $\kappa$ the distribution function has a Maxwellian core and a non-thermal power-law tail, while for large values of $\kappa$, the distribution is almost indistinguishable from a Maxwellian.}
\label{kappas}
\end{figure}

Examples of kappa distributions (\ref{kappa-dist}), for various values of the parameter $\kappa$, are shown in Figure~\ref{kappas}.  For high values of $\kappa$, the identity

\begin{equation}
e^{-x} = \lim_{\kappa \rightarrow \infty} \left ( 1 + {x \over \kappa} \right )^{-\kappa}
\end{equation}
shows that $f_\kappa(v)$ (Equation~(\ref{kappa-dist})) approaches the Maxwellian form

\begin{equation}\label{maxwellian}
f_\kappa(v) \sim \exp \left ( - {v^2 \over v_{\rm te}^2} \right ) \,\,\, .
\end{equation}

At low velocities $v \ll \sqrt{\kappa} \, v_{\rm te}$, the collisional diffusion term $f/\tau_{\rm d} \sim v^{-5}$ is dominant over the turbulent term $f/\tau_{\rm acc} \sim v^{-3}$, and so the distribution relaxes through collisional diffusion to a Maxwellian form.  Equation~(\ref{kappa-dist}) confirms that in this regime the kappa distribution approaches the form

\begin{equation}\label{low-velocity-limit}
f \sim  \left ( 1 - {v^2 \over v_{\rm te}^2} \right ) \quad {\rm as} \quad v \rightarrow 0 \,\,\, ,
\end{equation}
which is the same as the low-velocity limit of the Maxwellian distribution~(\ref{maxwellian}).  On the other hand, in the high-velocity limit $v \gg \sqrt{\kappa} \, v_{\rm te}$, the collisional diffusion timescale $\tau_{\rm d} \sim v^5$ (Equation~(\ref{tdiff-def})) is much longer than either the acceleration time $\tau_{\rm acc}$ (Equation~(\ref{tacc-def})) or the collisional deceleration timescale $\tau_{\rm c}$ (Equation~(\ref{tc-def})), both of which vary with velocity like $v^3$.  Thus in this regime the (temperature-dependent) collisional diffusion term is unimportant.  Further, since both $\tau_{\rm acc}$ and $\tau_{\rm c}$ have the same velocity dependence ($\sim v^3$), there is no characteristic velocity scale in this domain. Indeed, Equation~(\ref{kappa-dist}) confirms that the stationary distribution approaches a (scale-independent) power-law form:

\begin{equation}\label{kappa-power}
f_\kappa \rightarrow v^{-2 \kappa} \quad {\rm as} \quad v \rightarrow \infty \,\,\, .
\end{equation}

Overall, then, the use of the turbulent diffusion coefficient of the form~(\ref{dturb-result}) leads to an accelerated electron distribution that has the form of a kappa distribution~(\ref{kappa-dist}).  Such a distribution, as intended, accounts for the observed \citep[e.g.,][]{2003ApJ...595L..97H} blend of a Maxwellian core at low energies with a power law at higher energies. No artificial ``low-energy cutoff'' to the high-energy part of the distribution need be invoked; the electron distribution transitions smoothly from a ``non-thermal'' shape at high energies to a thermal (Maxwellian) form at low energies.

Recalling our remarks near the beginning of this Section on the possibility of a stationary Maxwellian form for $f(v)$ even in the presence of finite non-thermal turbulence, it should be noted that {\it any} scenario in which the turbulent acceleration time $\tau_{\rm acc}$ varies between $\tau_{\rm acc}\sim v^{5}$ at low velocities and $\tau_{\rm acc}\sim v^{3}$ at larger velocities will produce a kappa distribution.  Thus, we again note that the Maxwellian core of the kappa distribution is not necessarily associated with a collisionally-dominated thermal equilibrium state.

Now, the electron phase-space distribution function $f(v)$ is related to the mean electron flux ${\overline F}(E)$ (electrons~cm$^{-2}$~s$^{-1}$ per unit energy) through the relation $v f(v) \, d^3v = {\overline F}(E) \ dE$.  Using the elementary relation $E= m_e v^2/2$, it follows that $v \, d^3 v \sim v^3 \, dv \sim E \, dE$ and hence that $f(v) \sim {\overline F(E)}/E$.  Thus $\kappa$ is simply related to the power-law spectral index for the mean electron flux: ${\overline F}(E) \sim E^{-\delta}$, with $\delta = \kappa-1$.

Observations show that the hard X-ray spectrum above $\sim$(15-20)~keV is indeed approximately power-law in form: $I(\epsilon) \sim \epsilon^{-\gamma}$, with a typical value $\gamma \simeq$~5. The bremsstrahlung hard X-ray spectrum $I(\epsilon)$ (photons~cm$^{-2}$~s$^{-1}$~keV$^{-1}$ at the Earth) is related to the emitting mean electron flux spectrum ${\overline F}(E)$ (electrons~cm$^{-2}$~s$^{-1}$~keV$^{-1}$) \citep{2003ApJ...595L.115B} by

\begin{equation}\label{xray-electron}
I(\epsilon) = {n \, V \over 4 \pi R^2} \int_\epsilon^\infty {\overline F}(E) \, \sigma(\epsilon, E) \, dE \,\,\, ,
\end{equation}
where $V$ is the source volume, $R$ = 1 AU, and $\sigma(\epsilon,E)$ is the bremsstrahlung cross-section (cm$^2$~keV$^{-1}$), differential in photon energy $\epsilon$.  For the simple non-relativistic Kramers cross-section

\begin{equation}\label{kramers}
\sigma(\epsilon, E) \sim {1 \over \epsilon E} \,\,\, ,
\end{equation}
a hard X-ray spectrum $I(\epsilon) \sim \epsilon^{-\gamma}$ thus implies a mean electron flux spectrum ${\overline F}(E) \sim E^{-\delta}$, with $\delta = \gamma -1$; this relation also holds for more complex forms of $\sigma(\epsilon, E)$, such as the Bethe-Heitler cross-section \citep[see][]{1971SoPh...18..489B}. As discussed above, for the model considered here, the electron flux at high energies approximates a power-law with $\delta = \kappa-1$; thus the hard X-ray spectral index $\gamma$ and the electron distribution parameter $\kappa$ are equal:

\begin{equation}\label{kappa-gamma}
\kappa = \gamma \,\,\, ,
\end{equation}
and so a typical value of $\kappa \simeq 5$. Further, to obtain such a value of $\kappa$, Equation~(\ref{kappadef}) shows that the acceleration time

\begin{equation}\label{ratio-10}
{\tau_{\rm acc} } \simeq 10 \, \tau_{\rm c} \,\,\, ,
\end{equation}
i.e., about an order of magnitude larger than the collisional friction/deceleration time.

Since the power-law index $\kappa$ in this acceleration model is proportional to the acceleration time (Equation~(\ref{kappadef})), it follows that temporal hardening (softening) of the photon spectrum can be produced by a decrease (increase) of the acceleration time, resulting from a variation of the turbulent diffusion coefficient $D_0$ on a time-scale much longer than the overall relaxation time toward the steady state. The same argument was advanced by \citet{1977ApJ...211..270B} for interpreting spectral index variations of the photon spectrum during solar flares, including the commonly observed soft-hard-soft behavior.

\section{EVOLUTION TOWARD THE STATIONARY DISTRIBUTION}\label{evol}

We now consider in more detail the relaxation of the electron distribution function toward the stationary solution~(\ref{kappa-dist}), and in particular the formation of accelerated high-energy tails during such a process. This can be studied by introducing the function

\begin{equation}\label{udef}
u(v,t) \equiv \frac{f(v,t)}{f_\kappa (v)} \,\,\, .
\end{equation}
Substituting $f(v,t) = f_{\kappa}(v) \, u(v,t)$ into Equation~(\ref{fp-transformed}) and using the fact that $\partial f_\kappa /\partial t = 0$, we obtain an equation governing the evolution of the dimensionless quantity $u(v,t)$:

\begin{equation}\label{u-evol}
\frac{\partial u}{\partial t}=\frac{1}{v^{2}} \, \frac{\partial}{\partial v} \left ( v^{2} \, D(v) \, \frac{\partial u}{\partial v} \right )
-D(v) \, U^\prime(v) \, \frac{\partial u}{\partial v} \,\,\, .
\end{equation}
We now introduce the velocity-space variable $\eta$ through the transformation

\begin{equation}\label{yvtrans}
d\eta = \frac{dv}{\sqrt{D(v)}}
\end{equation}
and thus find that Equation~(\ref{u-evol}) can be written in the form of an advection-diffusion equation in velocity space:

\begin{equation}\label{fp-diffusion}
\frac{\partial u}{\partial t} + V(v) \, \frac{\partial u}{\partial \eta}=\frac{\partial^{2}u}{\partial \eta^{2}} \,\,\, .
\end{equation}
Here the advection speed (in velocity space) is given by

\begin{equation}\label{advection-speed}
V(v)=\sqrt{D(v)} \left [ U^\prime(v)-\frac{2}{v}-\frac{1}{2} \, {d \ln D(v) \over dv} \right ] \,\,\, .
\end{equation}

Because of the advection-diffusion structure of the Equation~(\ref{fp-diffusion}) that governs the relaxation toward the kappa distribution, the acceleration process is characterized by the successive energization of particles of higher and higher energy. The process may thus be described as a velocity-space ``front'' moving in the direction of increasing velocity
\citep{1957PhRv..107..350M, 1964pkt..book.....M, 2005PhRvE..72f1106C}.  The position $v_{f}(t)$ of this velocity-space front may be estimated by neglecting the diffusion term in Equation~(\ref{fp-diffusion}), so that

\begin{equation}\label{partialu-partialt}
\frac{\partial u}{\partial t} + V(v_{f}) \, \frac{\partial u}{\partial \eta} = 0 \,\,\, .
\end{equation}
The location of the velocity-space ``front'' may be identified with a fixed value of $u(v,t)=f(v,t)/f_\kappa$ (see Section~\ref{numerical}, where we set $u=0.5$).  Thus, setting the total derivative

\begin{equation}\label{du-total}
{du \over dt} \equiv \frac{\partial u}{\partial t} + {d \eta \over dt} \, \frac{\partial u}{\partial \eta} = 0
\end{equation}
allows us to write

\begin{equation}\label{vvf}
V(v_{f}) = \frac{d\eta}{dt} = \frac{1}{\sqrt{D(v_{f})}} \, \frac{dv_{f}}{dt}
\end{equation}
and hence

\begin{equation}\label{dvfdt}
\frac{dv_{f}}{dt} = \sqrt{D(v_{f})} \, V(v_{f}) = D(v_{f}) \,
\left [ U^\prime(v_{f})-\frac{2}{v_{\rm f}} - \frac{1}{2 \, v_{f}} \, {d \ln D(v_{f}) \over d \ln v_{f}} \right ] \,\,\, .
\end{equation}
In the high-velocity domain, $D(v) \simeq D_{\rm turb}(v) = D_0/v$ (Equations~(\ref{dv}) and~(\ref{dturb-result})) so that $d \ln D(v_f)/d \ln v_f = -1$.  Also, from Equation~(\ref{uprime-1-over-v}), in this regime $U^\prime(v_f) \simeq 2 \kappa /v_f$, so that Equation~(\ref{dvfdt}) reduces to

\begin{equation}\label{dvfdt-kappa}
\frac{dv_f}{dt} = {D(v_f) \over v_f} \, \left ( 2 \, \kappa - \frac{3}{2} \right ) =
\Gamma \left ( 1-\frac{3}{4\kappa} \right ) \, \frac{1}{v_f^{2}} \,\,\, ,
\end{equation}
where we have used Equations~(\ref{dturb-result}) and~(\ref{kappadef}).  This has solution

\begin{equation}\label{vf}
v_{f}(t) = \left ( 1 - \frac{3}{4\kappa}  \right )^{1/3} \, (3 \, \Gamma t)^{1/3} \, \simeq \, v_{\rm te} \left ( {t \over \tau} \right )^{1/3} \,\,\, ,
\end{equation}
where $\tau$ is the characteristic collision time for a thermal electron:

\begin{equation}\label{tau}
\tau = {v_{\rm te}^3 \over 3 \, \Gamma} =
{(2kT)^{3/2} \, m_e^{1/2} \over 12 \pi n e^4 \ln \Lambda}
\simeq \, 4 \times 10^{-3} \, {T^{3/2} \over n} \,\,\, .
\end{equation}
Substituting typical numerical values for the flaring corona in a dense looptop source, viz. $T= 2 \times 10^7$~K, $n=10^{11}$~cm$^{-3}$, we obtain $\tau \simeq 3 \, {\rm ms}$.  A hard-X-ray-producing electron has a typical energy $\sim$30~keV, about 15 times the thermal energy.  From Equation~(\ref{vf}) we see that

\begin{equation}
t = \tau \, \left ( {v_f \over v_{\rm te}} \right )^3 \,\,\, ,
\end{equation}
and hence the time to produce an electron of this energy is $\sim (15)^{3/2} \, \tau \simeq 0.2$~s, comparable to the observed rise and decay times of the hard X-ray flux at such energies.  This therefore raises the question of whether electrons can be confined in the acceleration region for a time sufficiently long for the ensemble to attain the asymptotic kappa distribution form~(\ref{kappa-dist}). We explore the consequences of this situation more fully in the following section.

\section{SPATIAL TRANSPORT AND ESCAPE}\label{escape-analysis}

In the acceleration model considered above, it is implicitly assumed that the electron distribution function is maintained close to isotropy as a result of efficient angular scattering in the acceleration region. This implies that the transport of electrons in this region is characterized by a spatial diffusion over length-scales much larger than their mean free-path $\lambda(v)$. Thus, after averaging over the fast pitch-angle scattering time-scale $\tau_{\rm pa}(v)\sim \lambda(v)/v$ responsible for isotropization of the distribution function, the pitch-angle-dependent streaming transport of electrons parallel to the background magnetic field, described by the relation

\begin{equation}
\dot{z}=\mu \, v
\end{equation}
(where $z$ is the coordinate along a direction parallel to the guiding magnetic field and $\mu$ is the cosine of the angle between the velocity and magnetic field vectors), assumes the diffusive form

\begin{equation}\label{kappa-par}
\mu \, v \, \frac{\partial f}{\partial z} \rightarrow \frac{\partial}{\partial z} \,
\left [ K_{\parallel} \, \frac{\partial f}{\partial z} \right ] \,\,\, .
\end{equation}
where the corresponding spatial diffusion coefficient is given by

\begin{equation}
K_\parallel = \frac{\lambda (v) \, v}{3} \,\,\, .
\end{equation}
In this strong scattering limit \citep[e.g.,][]{2012SSRv..173..535P}, the Fokker-Planck equation, including the spatial transport term, takes the form

\begin{equation}\label{fp-kappa-parallel}
\frac{\partial f}{\partial t} +  \frac{\partial}{\partial z} \, \left [ {\lambda(v) \, v \over 3} \,
\frac{\partial f}{\partial z} \right ] =\frac{1}{v^{2}} \, \frac{\partial }{\partial v} \,
\left \{ v^{2} \left [ \left ( \frac{\Gamma \, v_{\rm te}^{2}}{2v^{3}}+D_{\rm turb}(v) \right ) \,
\frac{\partial f}{\partial v}+\frac{\Gamma}{v^{2}} \, f \right ] \right \} \,\,\, .
\end{equation}
Now, representing $\partial/\partial z$ as $1/L$, thus defining the ``length'' $L$ of the acceleration region, Equation~(\ref{fp-kappa-parallel}) can be written

\begin{equation}\label{escape}
\frac{\partial f}{\partial t}=\frac{1}{v^{2}} \, \frac{\partial }{\partial v} \,
\left \{ v^{2} \left [ \left ( \frac{\Gamma \, v_{\rm te}^{2}}{2v^{3}}+D_{\rm turb}(v) \right ) \,
\frac{\partial f}{\partial v}+\frac{\Gamma}{v^{2}} \, f \right ] \right \}-\frac{f}{\tau_{\rm esc}(v)} \,\,\, ,
\end{equation}
where the escape time-scale

\begin{equation}\label{tau-esc-v}
\tau_{\rm esc}(v) = {3 L^2 \over \lambda(v) \, v} = \left ( {3L \over \lambda(v)} \right ) \, \left ( \frac{L}{v} \right ) \,\,\, .
\end{equation}

In this leaky-box approximation, intended to represent the effect of spatial transport out of the acceleration region, the role of the escape term is to deplete the number of electrons from the acceleration region over a transport time scale $\tau_{\rm esc}(v)$. We notice that this diffusive escape time becomes of the order of the free-streaming escape time $L/v$ only when the mean free path $\lambda$ and the acceleration region length $L$ are comparable.  We also note that in the absence of an additional source of particles maintaining a steady state, the number of electrons will decrease with time as a result of the escape term.

\begin{figure}[htpb]
\centering
\includegraphics[width=10cm]{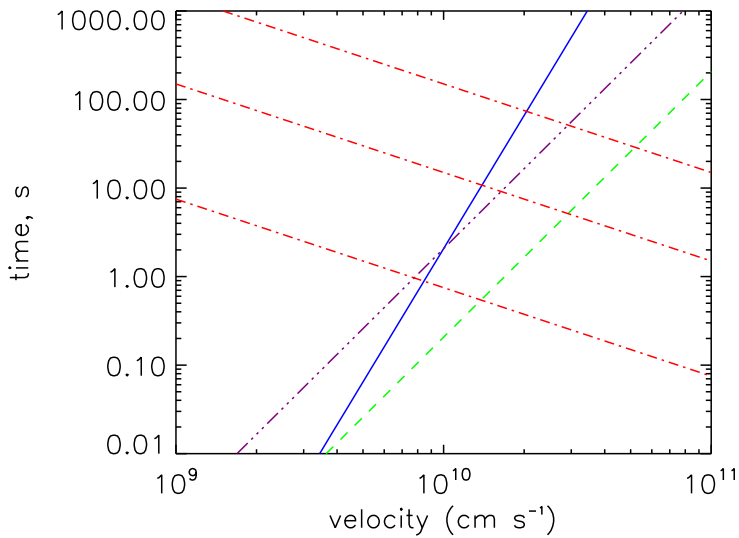}
\caption{Characteristic timescales of the system. The solid blue line represents the collisional diffusion timescale $\tau_{\rm d} \propto v^5$ (Equation~(\ref{tdiff-def})), the purple triple-dot-dash line the acceleration timescale $\tau_{\rm acc} \propto v^3$ (Equations~(\ref{tacc-def}) and~(\ref{dturb-result})), the green dashed line the collisional deceleration timescale $\tau_{\rm c} \propto v^3$ (Equation~(\ref{tc-def})), and the red dot-dashed lines the escape time $\tau_{\rm esc} \propto v^{-1}$ (Equation~(\ref{tau-esc-v})) for (from bottom to top) $\lambda/L = 0.2$, $0.01$ and $0.001$. } \label{char-times-fig}
\end{figure}

In equation (\ref{escape}), there are now four terms (acceleration, collisional deceleration, collisional diffusion, and escape), each with their associated characteristic timescale. The relative importance of these terms is summarized on Figure~2.  Ignoring for the moment the (red dash-dot) lines representing the escape time $\tau_{\rm esc}(v)$ ($\sim v^{-1}$ for a velocity-independent mean free path $\lambda$), we can see the two regimes that define the boundaries of the kappa distribution. At low velocities the collisional diffusion time, $\tau_{\rm d}(v) \sim v^5$ (blue line), is shorter than, and hence dominant over, the acceleration timescale  $\tau_{\rm acc}(v) \sim v^3$ (purple triple-dot-dash line); this creates a collisionally-dominated Maxwellian core. At higher velocities, the physics is dominated by the acceleration and collisional friction timescales $\tau_{\rm acc}(v)$ and $\tau_{\rm c}(v)$, which have the same velocity dependence $\sim v^3$.  The resulting absence of characteristic velocity in this regime yields a power-law spectrum with index $\kappa=\tau_{\rm acc}/2\tau_{\rm c}$.

Since the mean free path $\lambda(v)$ can generally be expected to be constant or increase with $v$, the escape time scale $\tau_{\rm esc}(v)$ will also generally be a decreasing function of $v$.  Thus, at sufficiently large velocities, $\tau_{\rm esc}$ will eventually become smaller than all of $\tau_{\rm d}(v)$, $\tau_{\rm acc}(v)$, and $\tau_{\rm c}(v)$ (all of which are increasing functions of $v$). Hence we define the escape velocity $v_{\rm esc}$ as the critical velocity where escape starts to be the leading effect, found by equating $\tau_{\rm esc}(v)$ (Equation~(\ref{tau-esc-v})) and $\tau_{\rm c}(v)$ (Equation~(\ref{tc-def})):

\begin{equation}
{3 L^2 \over \lambda(v_{\rm esc}) \, v_{\rm esc} } = \frac{v_{\rm esc}^3} {\Gamma} \,\,\, .
\end{equation}
For $\lambda = \lambda_0 (v/v_0)^{-\alpha}$ (see discussion in Section~\ref{stationary}), the explicit solution is

\begin{equation}
v_{\rm esc} = \left ( {3 L^2 \, \Gamma \over \lambda_0 \, v_0^\alpha} \right )^{1 \over 4 - \alpha} \,\,\, .
\end{equation}

The three red dot-dashed lines in Figure~\ref{char-times-fig} show $\tau_{\rm esc}(v)$ for three different values of $\lambda/L$ (where $\lambda$ is assumed to be independent of velocity, i.e. $\alpha = 0$). As the mean free path decreases the escape time becomes longer (Equation~(\ref{tau-esc-v})) and thus the intersection with $\tau_{\rm c}(v)$ occurs at a higher velocity. For example, for $\lambda/L = 0.2$, $\tau_{\rm esc}(v)$ intercepts $\tau_{\rm c}(v)$ before the acceleration timescale has become shorter than the collisional diffusion timescale, $\tau_{\rm d}(v)$, so in this case we don't expect a kappa distribution to form. On the other hand, looking at the $\lambda/L = 0.0001$ line we see that the escape time intercepts at much larger velocities, so that a kappa distribution power law tail will form for $v \ll v_{\rm esc}$, the shape of the distribution function becoming substantially different from a kappa distribution only at large velocities $v\gapprox v_{\rm esc}$.

We may solve the ``leaky box'' Fokker-Planck equation (\ref{escape}) using an approximation based on an analogy to the pitch-angle loss-cone in a magnetic trap where in the loss-cone situation, there is a critical pitch angle below which electrons escape and above which they remain fully trapped.  By analogy, the ``escape velocity'' $v_{\rm esc}$ is the velocity below which electrons are considered to remain in the acceleration region and above which they are considered to freely escape\footnote{this approximation is also used to model the escape of stars from gravitational clusters \citep{1943ApJ....97..263C,1958ApJ...127..544S}}. The Fokker-Planck equation may therefore be written without an explicit escape term:

\begin{equation}\label{escc}
\frac{\partial f}{\partial t}=\frac{1}{v^{2}} \, \frac{\partial }{\partial v} \,
\left \{ v^{2} \left [ \left ( \frac{\Gamma \, v_{te}^{2}}{2v^{3}}+D_{\rm turb}(v) \right ) \,
\frac{\partial f}{\partial v}+\frac{\Gamma}{v^{2}} \, f \right ] \right \} \,\,\, ,
\end{equation}
together with the absorbing boundary condition

\begin{equation}\label{boundary}
f(v_{\rm esc},t)=0
\end{equation}
replacing the escape term.

We treat the problem in the limit of a small escape rate when the acceleration region can effectively be considered a collisional thick target.  Thus the time-dependent solution can be found by perturbation analysis.  We start with the Fokker-Planck equation in the form~(\ref{fp-transformed}), repeated here:

\begin{equation}\label{fokker-repeated}
\frac{\partial f}{\partial t} = \frac{1}{{v^{2}}} \, \frac{\partial}{\partial v} \,
\left [ v^{2} \, D(v) \left ( \frac{\partial f}{\partial v} + f \, U^\prime(v) \right ) \right ] \,\,\, ,
\end{equation}
which is to be solved subject to the boundary condition~(\ref{boundary}). We posit a solution of the form

\begin{equation}\label{eigenmode-sol}
f(v,t)=A \, e^{\nu t} \, g(v) \,\,\, ,
\end{equation}
leading to

\begin{equation}\label{eigenvalue-equation}
\nu \, g(v) =\frac{1}{{v^{2}}} \, \frac{\partial }{\partial v} \,
\left [ v^{2}D(v) \left ( \frac{dg(v)}{dv} +g(v) \, \frac{dU}{dv} \right ) \right ] \,\,\, .
\end{equation}
This has a first integral

\begin{equation}\label{first-integral}
\frac{dg(v)}{dv} + g(v) \, \frac{dU(v)}{dv} = \frac{\nu}{v^{2} D(v)} \, \int_{0}^{v}dw \, w^{2} \, g(w) \,\,\, .
\end{equation}
We next write the solution as an expansion in the decay rate $\nu$
\citep[cf.][]{1965AJ.....70..376K,2010PhyA..389.1021L}:

\begin{equation}\label{expamsion}
g(v)=g_{0}(v)+\nu \, g_{1}(v) + \ldots \,\,\, .
\end{equation}
The zero-order equation is

\begin{equation}\label{zero-order}
\frac{dg_{0}(v)}{dv}+g_{0}(v) \, \frac{dU(v)}{dv} = 0 \,\,\, ,
\end{equation}
with the expected solution (see Equation~(\ref{ss}))

\begin{equation}\label{zero-order-soluton}
g_{0}(v) = A \, e^{-U(v)} \,\,\, ,
\end{equation}
where $A$ is a normalization factor.  The first-order equation is

\begin{equation}\label{first-order}
\frac{dg_{1}(v)}{dv} + g_{1}(v) \, \frac{dU(v)}{dv} = \frac{1}{v^{2}D(v)}\int_{0}^{v} dw \, w^{2} \, g_{0}(w) =
\frac{A}{v^{2}D(v)}\int_{0}^{v} dw \, w^{2} \, e^{-U(w)} \,\,\, .
\end{equation}
Using an integrating factor $e^{U(v)}$, we derive the solution

\begin{equation}\label{first-order-solution}
g_{1}(v) =  A \, e^{-U(v)} \, \chi(v) \,\,\, ,
\end{equation}
where $\chi(v)$ is the function defined by

\begin{equation}\label{chi-def}
\chi^\prime (v) = \frac{e^{U(v)}}{v^{2}D(v)} \int_{0}^{v} dw \, w^{2} \, e^{-U(w)} \,\,\, .
\end{equation}
Using Equations~(\ref{expamsion}), (\ref{zero-order-soluton}), and~(\ref{first-order-solution}), the distribution function is, to first order in $\nu$, given by

\begin{equation}\label{time-dept-soln}
f(v,t) = A \, e^{-U(v)} \, e^{\nu t} \, [ \, 1 + \nu \, \chi(v) \, ] \,\,\, .
\end{equation}
Now introducing the boundary condition $f(v_{\rm esc},t)=0$ (Equation~(\ref{boundary})), we obtain the identification

\begin{equation}\label{nu-time-soln}
\nu = - \, \frac{1}{\chi(v_{\rm esc})} \,\,\, ,
\end{equation}
which is the sought-after escape rate in the limit of large escape velocity.

In the case where $D_{\rm turb}(v)=D_{0}/v$ (Equation~(\ref{dturb-result})), we recall that (cf. Equation~(\ref{dv}))

\begin{equation}\label{dv-time-soln}
D(v)=\frac{\Gamma \, v_{\rm te}^{2}}{2 \, v^{3}}+\frac{D_{0}}{v} \,\,\, ,
\end{equation}
and that (Equation~(\ref{u-solution}))

\begin{equation}\label{uv-time-soln}
U(v)=\kappa \ln \left ( 1 + { v^2 \over \kappa v_{\rm te}^2} \right ) \,\,\, .
\end{equation}
Substituting results~(\ref{nu-time-soln}) and~(\ref{uv-time-soln}) in Equation~(\ref{time-dept-soln}) gives the (normalized; see Equation~(\ref{kappa-dist})) time-dependent solution of the leaky-box acceleration model:

\begin{equation}\label{leaky-box-soln}
f(v,t)=
\frac{n \, e^{-t/\chi(v_{\rm esc})}}{\pi^{3/2} \, v_{te}^{3} \, \kappa^{3/2}} \, \frac{\Gamma(\kappa)}{\Gamma \left (\kappa-\frac{3}{2} \right ) } \, \left ( 1+\frac{v^{2}}{\kappa \, v_{\rm te}^{2}} \right )^{-\kappa} \left [ 1-\frac{\chi(v)}{\chi(v_{\rm esc})} \right ] \,\,\, .
\end{equation}
The last factor in brackets describes the deviation from the kappa distribution and

\begin{equation}
n(t)=n \, \exp[{-t/\chi(v_{\rm esc})}]
\end{equation}
describes the decreasing overall number of particles in the box with time, which are both a consequence of escape of particles out of the acceleration region. These functions depend on the function $\chi(v)$, which is determined through Equations~(\ref{chi-def}), (\ref{dv-time-soln}), and~(\ref{uv-time-soln}):

\begin{equation}\label{chi-expression}
\chi^\prime(v) = {v \over \kappa \, D_0 \, v_{\rm te}^2} \left ( 1 + \frac{v^{2}}{\kappa \, v_{\rm te}^{2}} \right )^{\kappa-1}
\, \int_{0}^{v} dw \, w^{2} \left ( 1+\frac{w^{2}}{\kappa \, v_{\rm te}^{2}} \right )^{-\kappa} \,\,\, .
\end{equation}
As a reminder, the above solution is valid in the limit where the acceleration region behaves essentially as a thick target. A stationary solution of a similar leaky-box Fokker-Planck equation, without the collisional diffusion term but with a source of particles, was also obtained by \citet{1977ApJ...211..270B}.

\section{NUMERICAL SOLUTIONS}\label{numerical}

\begin{figure}[htpb]
\centering
\includegraphics[width=10cm]{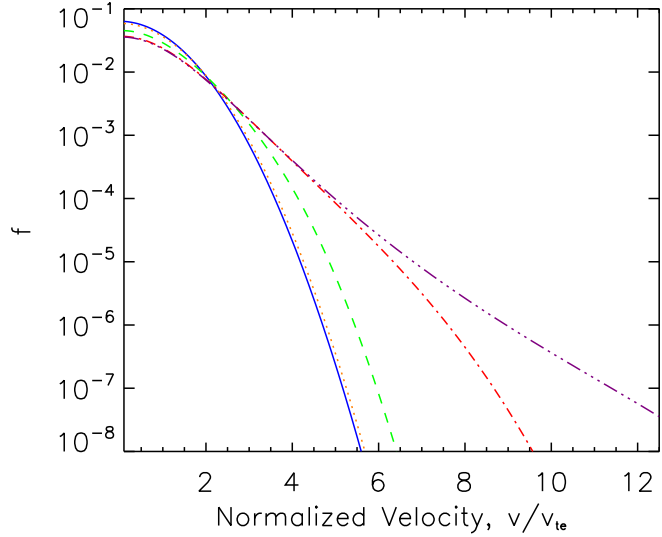}
\caption{Temporal evolution of electron distribution function $f(v,t)$, for $\kappa \, (\equiv \Gamma/2D_0) = 5$.  The solid blue line shows the initial Maxwellian and then, from left to right, $f(v,t)$ at $t/\tau_c = 1.0$ (orange dotted line),  $t/\tau_c =10$ (green dashed line), $t/\tau_c=100$ (red dot-dashed line), and $t/\tau_c=1000$ (purple dot-dot-dot-dashed line). }\label{dist-evol-fig}
\end{figure}

We have performed a number of numerical solutions of the Fokker-Planck equation~(\ref{fund}), with the goal of validating the analytical approximations of Section~\ref{evol}.  We use a finite difference code to examine the evolution of the electron velocity distribution $f(v,t)$ with time as governed by Equation~(\ref{fund}) with $D_{\rm turb}=D_0/v={\Gamma}/2\kappa v$. For the simulations, we adopted a typical value for $\kappa =5$, which agrees well with solar flare hard X-ray observations (cf. Equation~(\ref{kappa-gamma})).

Firstly we check that we do indeed obtain a kappa distribution from the balance of Coulomb collisions and stochastic acceleration within Equation~(\ref{fund}). Figure~\ref{dist-evol-fig} shows the evolution of an originally Maxwellian thermal population of electrons (blue, solid line) toward a final state which agrees with the stationary solution kappa distribution (purple, dot-dot-dot-dashed line) as given by Equation~(\ref{kappa-dist}).  We see that the distribution at $t = 100 \, \tau_{\rm c}$ closely approximates the kappa distribution form below $\simeq 5 \, v_{\rm te}$, corresponding to a range of about four orders of magnitude in $f(v,t)$. Such a distribution at $t=100 \, \tau_{\rm c}$ is thus close to a kappa distribution form for a significant proportion of the particles.

\begin{figure}[htpb]
\centering
\includegraphics[width=10cm]{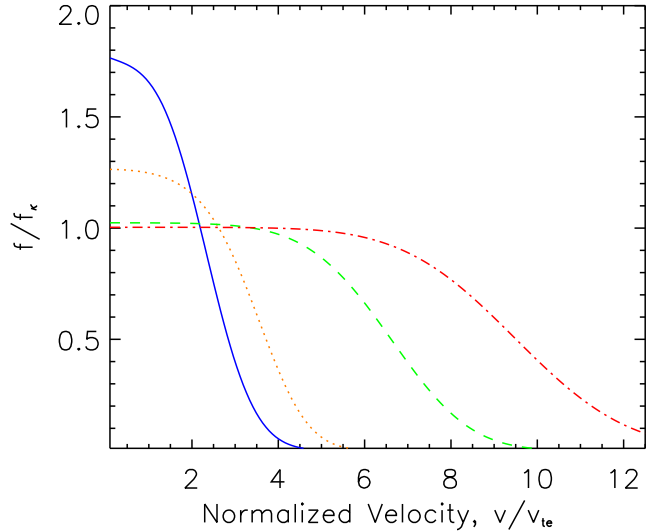}
\caption{Evolution of the normalized distribution $f/f_{\kappa}$ with time. The solid blue line shows the normalized injected Maxwellian and then, from left to right: $f(v,t)/f_{\kappa}$ at $t/\tau_c = 10$ (orange dotted line), $t/\tau_{\rm c} =100$ (green dashed line), and $t/\tau_{\rm c}=300$ (red dot-dashed line). }\label{norm-dist-evol-fig}
\end{figure}

The evolution of the normalized distribution (Figure~\ref{norm-dist-evol-fig}) shows a ``wavefront'' moving towards higher energies, as expected from the advection-diffusion nature of Equation~(\ref{u-evol}).  (We do not plot the final state of the distribution at $t = 1000 \, \tau_{\rm c}$ as it is almost a constant across the domain.)  Examining $f/f_\kappa$ gives a clearer view of how close the electron distribution approximates a kappa distribution at different points of the simulation.  Our results confirm that the electron distribution function at $t = 100 \, \tau_{\rm c}$ (green dashed line) is close to a kappa distribution for $v \lapprox  5 \, v_{\rm te}$ and that for $t = 300 \, \tau_{\rm c}$ it is almost indistinguishable from a kappa distribution up to around $7 \, v_{\rm te}$.

\begin{figure}[htpb]
\centering
\includegraphics[width=10cm]{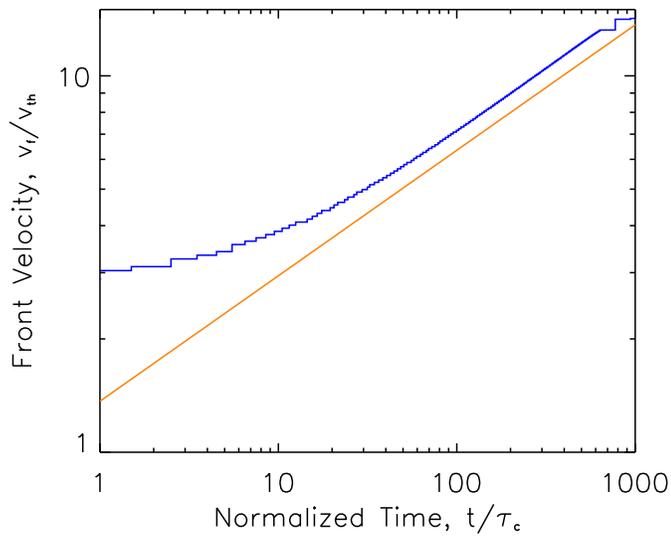}
\caption{Location $v_f$ of the front in velocity space (in units of the thermal speed $v_{\rm te}$) versus time (in units of the collisional deceleration time $\tau_{\rm c}$). The analytic approximation for front speed $v_f(t)$ (Equation~(\ref{vf})) is shown by the orange solid line. The blue line shows the location of the velocity where $f/f_{\kappa}=0.5$, from numerical simulations.} \label{vf-comparison-fig}
\end{figure}

In Section~\ref{evol} we found that the location of the front in velocity space evident in Figure~\ref{norm-dist-evol-fig} should depend on time according to $v_f(t) \sim t^{1/3}$ (Equation~(\ref{vf})). To assess the accuracy of this analytical result we arbitrarily choose a value $u(v,t) = f(v,t)/f_\kappa = 0.5$ to define the front location $v_f(t)$. A plot of $v_f$ versus time (in units of the collision time $\tau_{\rm c}$) is shown in Figure~\ref{vf-comparison-fig}.  Before $t \simeq 20 \, \tau_{\rm c}$ there is a significant disagreement because the analytic expression~(\ref{vf}) holds only for $t \gg \tau_{\rm c}$, i.e., when a sufficient number of particles have been accelerated to non-thermal energies.  At longer times $t \gapprox 700 \, \tau_{\rm c}$ a discrepancy also develops, which is due to the simulation results reaching the upper limit of velocity allowed in the system. However, for times between these two extremes, we see excellent agreement between the numerical and analytic solutions in terms\footnote{The constant offset between the curves in Figure~\ref{vf-comparison-fig} is not significant; it merely reflects the subjective nature of the choice $f(v,t)/f_{\kappa} = 0.5$ for the location of the velocity front.} of the power-law slope $d \ln v_f/d \ln t =1/3$. These numerical results show that the velocity-space front scenario as well as Equation~(\ref{vf}) provide a generally good description of the way particles are accelerated toward the kappa distribution in this model.

\section{STOCHASTIC ACCELERATION BY A LARGE SCALE ELECTRIC FIELD WITH STRONG PITCH-ANGLE SCATTERING}\label{stationary}

The primary energy release in solar flares involves the reconnection of magnetic fields to produce electric fields. Various authors have considered the role of magnetic reconnection in particle acceleration, including large-scale sub-Dreicer \citep[e.g.,][]{1994ApJ...435..469B} and supra-Dreicer \citep[e.g.,][]{1993SoPh..146..127L,1996ApJ...462..997L} electric fields. Here we extend the analysis of large-scale coherent electric fields to include the role of turbulent pitch-angle scattering. A main objective of this analysis is to point out that efficient pitch-angle scattering of the particles in a region of constant electric field strength can still create an effect akin to stochastic acceleration, possibly suppressing the runaway phenomenon \citep{1994ApJ...435..469B} and preventing the production of an unacceptably large unidirectional current in these acceleration models. As we are interested in the formation of kappa distributions by turbulent acceleration we also discuss the conditions leading to a turbulent diffusion coefficient of the desired form $D_{\rm turb}(v)\sim v^{-1}$.

Under the action of an accelerating electric field $E_\parallel$ (statvolt~cm$^{-1}$) parallel to the ambient magnetic field ${\bf B}$, the one-dimensional kinetic equation for a gyrotropic ($\partial f/\partial \phi=0$) distribution function $f(z,\beta, v,t)$ is

\begin{equation}\label{kinetic}
\frac{\partial f}{\partial t} + v \, \cos \beta \, \frac{\partial f}{\partial z} + \frac{eE_\parallel}{m_{e}} \, \mathbf{b}.\nabla_{\mathbf{v}}f =
\frac{v}{\lambda} \, \frac{1}{\sin \beta} \, \frac{\partial}{\partial \beta} \left ( \sin \beta \, \frac{\partial f}{\partial \beta} \right ) \,\,\, ,
\end{equation}
where $z$ (cm) is the position of the gyrocenter along the magnetic field with direction ${\mathbf b} = {\mathbf B_0}/B_0$, $\beta$ is the pitch angle ($\cos \beta = \mathbf {v}.\mathbf{B}_{0}/vB_{0} = v_{\parallel}/v$) and $v=\sqrt{v_{\parallel}^{2}+v_{\perp}^{2}}$ is the particle speed. In the case under consideration, the acceleration region is characterized by an electric field of constant magnitude $E_{\parallel}$ aligned with the direction $\mathbf{b}$ of the magnetic field. Transforming to the variables $(z,\mu,v,t)$, with $\mu= \cos \beta$, this may be rewritten as

\begin{equation}\label{fokker}
\frac{\partial f}{\partial t}+\mu \, v \, \frac{\partial f}{\partial z} +
\frac{e E_\parallel}{m_{e}} \, \mu\, \frac{\partial f}{\partial v} +\frac{e E_\parallel}{m_{e}} \, \frac{(1-\mu^{2})}{v}
\, \frac{\partial f}{\partial \mu} = \frac{v}{\lambda} \, \frac{\partial}{\partial \mu} \left [ (1-\mu^{2}) \, \frac{\partial f}{\partial \mu} \right ] \,\,\, .
\end{equation}
The last term in this equation describes pitch-angle diffusion, which tends to isotropize the distribution function on a (velocity-dependent) time scale given by

\begin{equation}\label{taupa-def}
\tau_{\rm pa}(v)=\frac{\lambda(v)}{v} \,\,\, ,
\end{equation}
where the mean free path $\lambda(v)$ is generally a function of $v$. For instance, collisional pitch-angle scattering produces isotropization of the distribution function on time scale $\tau_{\rm pa}\sim v^{3}$ ($\sim v_{\rm te}^{3}$) corresponding to $\lambda\sim v^{4}$ ($\sim T^{2}$ for thermal particles) and the absence of an external electric field\footnote{for $\lambda \sim v^{4}$, Equation~(\ref{fokker}) is identical to the model studied by \citet{1964PhFl....7..407K} in the context of the formation of runaway electrons in plasmas, while the case of a velocity-independent mean free-path $\lambda \sim v^{0}$ corresponds to the standard Drude model of electric resistivity studied by Lorentz (1905).}.

Acceleration of the particles is described by the third term in Equation~($\ref{fokker}$), i.e.,

\begin{equation}\label{dot-v}
\dot{v}=\frac{e E_\parallel}{m_{e}} \, \mu \,\,\, ,
\end{equation}
which shows that fluctuations in $\mu$ are also responsible for fluctuations in $v$. Given that particles are accelerated by the external electric field, it is quite clear that the isotropization effect of pitch angle scattering becomes dominant whenever $\tau_{\rm pa}(v) = \lambda(v)/v$ is a decreasing function of $v$. In fact, assuming $\tau_{\rm pa}(v)\sim v^{-\alpha}$, it can be shown that when $\alpha>0$ the particle distribution function remains close to isotropic despite the presence of the constant electric force \citep{1981JSP....24...45P,PhysRevE.56.3822,PhysRevLett.99.030601}. No runaway phenomenon occurs in  this case.  The combined effect of the electric field and pitch-angle scattering shows up as isotropic diffusive acceleration of electrons and an unlimited growth of their kinetic energy in the absence of collisional energy losses. The corresponding velocity-space diffusion coefficient can then be computed from the \citet{taylor} formula

\begin{equation}\label{taylor-dturb}
D_{\rm turb}(v) = \frac{e^{2}E_{\parallel}^{2}}{m_{e}^{2}} \int_{0}^{\infty} \langle \, \mu(0) \, \mu(t) \, \rangle  \, dt
=\frac{e^{2} \, E_\parallel^{2} \, \lambda(v)}{3 \, m_{e}^{2} \, v} \,\,\, .
\end{equation}
In the case where the mean free path $\lambda(v)$ is independent of $v$, this may be written

\begin{equation}\label{taylor-dturb-one-over-v}
D_{\rm turb}(v) = {D_{0} \over v} \,\,\, ,
\end{equation}
with

\begin{equation}\label{D0-result}
D_{0} = \frac{e^{2} \, E_{\parallel}^{2} \, \lambda}{3 \, m_{e}^{2}} \,\,\, .
\end{equation}
As discussed in Section~\ref{kap}, in such a case the acceleration time $\tau_{\rm acc}(v) \sim v^2/D_{\rm turb}(v)$ (Equation~(\ref{tacc-def})) and the collisional deceleration time $\tau_{\rm c}(v)$ (Equation~(\ref{tdiff-def})) have the {\it same velocity dependence} ($\sim v^3$). Indeed, since $D_{\rm turb}(v) \propto \tau_{\rm pa}(v) = \lambda/v \sim v^{-1}$, Equation~(\ref{tacc-def}) shows that $dv/dt \sim v/\tau_{\rm acc} \sim v^{-2}$, so that $dE/dt \equiv m_e \, v \, dv/dt \sim v^{-1} \sim E^{-1/2}$; the kinetic energy thus grows like $E\propto t^{2/3}$ (see Equation~(\ref{vf})). The role of collisional energy losses is to allow the distribution of electrons to steadily converge toward the stationary kappa distribution~(\ref{kappa-dist})
as a result of the balance between turbulent acceleration and friction.

In the above reasoning we have completely ignored the finite size of the acceleration region which imposes a maximum energy gain bounded by the finite electric potential drop across the acceleration region. Unfortunately, our treatment (Section~\ref{escape-analysis}) of escape from a finite-length acceleration region does not apply to the case of a stationary electric field -- whereas the maximum energy gained by particles from time-dependent electric fields in a finite length acceleration region depends on the confinement (escape) time $\tau_{\rm esc}$, the amount of energy gained by particles under the influence of a time-independent electric field is independent of the amount of time these particles stay confined in the acceleration region.  Further, given that only particles moving parallel to the applied electromotive force $eE_\parallel$ gain energy, while those flowing antiparallel to the applied force lose it, a spatial asymmetry remains in such an acceleration model, despite the effects of isotropization. This is an undesirable feature of the model, requiring that some form of fragmentation of the electric field, such as oppositely directed electric fields on different magnetic field lines \citep[e.g.,][]{1985ApJ...293..584H,1995ApJ...446..371E,1997ApJ...489..367A,2004ApJ...608..540V,2008ApJ...687L.111B,2012SoPh..277..299G,2012SSRv..173..223C,2013SoPh..284..489G}, must be invoked.

The value of $\kappa$ in the distribution~(\ref{kappa-dist}) is the ratio (Equation~(\ref{kappadef})) of the collision parameter $\Gamma$ (Equation~(\ref{gamma-def})) to the diffusion parameter $D_0$ (Equation~(\ref{dturb-result})) and hence, in a model involving stochastic acceleration by direct electric field, relates the ambient density $n$ to the (square of the) strength of the accelerating electric field $E_\parallel$ (Equation~(\ref{D0-result})).  Thus the shape of the accelerated electron distribution constrains the values of one or both of these physical parameters.  We now briefly explore the nature of this constraint as imposed by the observed shape of solar flare hard X-ray spectra.

Using Equations~(\ref{kappadef}) and (\ref{D0-result}), we find that the value of the power-law index $\kappa$ in the distribution~(\ref{kappa-dist}) is given by

\begin{equation}\label{kappa-dreicer}
\kappa = \frac{\Gamma}{2D_{0}}=\frac{3}{2} \left ( \frac{\lambda _{c}}{\lambda} \right ) \left ( \frac{E_{D}}{E_{\parallel}} \right )^{2} \,\,\, ,
\end{equation}
where we have introduced the usual collisional mean free-path

\begin{equation}\label{lambda-coll}
\lambda_{c} = \frac{(k_B T)^{2}}{4\pi ne^{4}\ln \Lambda}
\end{equation}
and the Dreicer field

\begin{equation}\label{dreicer-def}
E_D \equiv {k_B T \over e \lambda_c} = {4 \pi n \, e^3 \ln \Lambda \over k_B T} \,\,\, ,
\end{equation}
i.e., the field strength required to accelerate an electron to the thermal energy over a distance equal to the collisional mean free path.  As discussed in Section~\ref{kap}, observations of solar flare hard X-ray spectral shapes reveal that a typical value for $\kappa$ is $\kappa \simeq 5$ (Equation~(\ref{kappa-gamma})), which therefore provides the following constraint on the value of the accelerating electric field:

\begin{equation}\label{e-constraint}
E_{\parallel} \simeq \left ( \frac{3}{10}\frac{\lambda_{c}}{\lambda} \right )^{1/2} \, E_{D} \,\,\, .
\end{equation}
Moreover, normalizability of the kappa distribution~(\ref{kappa-dist}) requires that $\kappa > 3/2$, or $
E_{\parallel}< \left ( {\lambda_{\rm c} / \lambda } \right ) ^{1/2}E_{D}$. For typical conditions in the flaring loop-top source coronal plasma, $T \simeq 2 \times 10^7$~K and $n \simeq 10^{11}$~cm$^{-3}$, leading to a collisional mean free path $\lambda_c \simeq 5 \times 10^6$~cm and a Dreicer field $E_D \simeq 3 \times 10^{-4}$~V~cm$^{-1}$.  Recently \citet{2014ApJ...780..176K} have argued, on the basis of the observed variation of hard X-ray source size with energy, that the turbulent mean free path $\lambda$ is in the range $10^8 - 10^9$~cm. Thus $\lambda_c/\lambda \simeq 0.005 - 0.05$, leading (Equation~(\ref{e-constraint})) to $E_{\parallel} \simeq (0.05 - 0.1)E_D \simeq (2 - 3) \times 10^{-5}$~V~cm$^{-1}$. Such a value of $E_{\parallel}$ is broadly consistent with the acceleration of electrons to deka-keV energies over observed loop lengths $L \simeq 10^9$~cm.

\section{SUMMARY AND CONCLUSIONS}\label{conclusion}

Driven by {\em RHESSI} observations of confined loop-top hard X-ray sources in solar flares, we have considered a model with cospatial stochastic acceleration, collisional deceleration and thermalization, and hard X-ray bremsstrahlung emission.  For a turbulent diffusion coefficient associated with the acceleration mechanism of the form $D_{\rm turb} \sim 1/v$, and in the absence of particle escape, the electron distribution asymptotically approaches a kappa distribution~(\ref{kappa-dist}) with time.

The approach toward this asymptotic steady-state kappa distribution proceeds as a ``wavefront'' in velocity space, with electrons of speed $v$ accelerated at successively greater times $t \sim v^3 \sim E^{3/2}$. This velocity-space front scenario, as well as the basic timescales involved, are supported by the results of numerical simulations. For sufficiently high velocities, the time taken to approach the kappa distribution becomes long enough that escape of electrons from the acceleration region can no longer be neglected.  The effect of this was considered analytically in the limit of a small escape rate, when the acceleration region effectively behaves as a thick-target.

With the high-spectral-resolution hard X-ray observations from {\em RHESSI}, the form of the hard X-ray-emitting (and hence, in this context, accelerated) electron distribution can be determined with impressive accuracy.  Analysis of the spatially-integrated spectra from loop-top sources therefore provides a test of the predictions of the current model.  Further, quantitative analysis of the energies, both low and high, at which the inferred electron distribution approaches and/or deviates from the asymptotic, escape-free, kappa distribution provides information on the value of the physical parameters of the model, such as the acceleration region length $L$ and the diffusion coefficient parameter $D_0$.

\acknowledgments We thank the referee for drawing our attention to the different representations of the kappa distribution and its broader role in space plasma physics. This work is partially supported by a STFC grant. Financial support  by the European Commission through the ``Radiosun'' (PEOPLE-2011-IRSES-295272) is gratefully acknowledged. AGE was supported by grant number NNX10AT78G from NASA's Heliospheric Physics Division.

\bibliographystyle{apj}
\bibliography{kappa}

\end{document}